\documentclass[aps,prd,nofootinbib,showpacs,preprintnumbers]{revtex4}

\usepackage[dvipdfm]{graphicx}

\usepackage{color}
\usepackage{amsmath,amssymb}	% required for `\align' (yatex added)

\begin{document}
%%%%%%%%%%%%%%%%%%%%%%%%%%%%%%%%%%%%%%%%%%%%%%%%%%%%%%%%%%%%%%%%%%%%%%%%%%%%%%%

\title{
LHC Run-I constraint on the mass of doubly charged Higgs bosons \\
in the same-sign diboson decay scenario
}

\author{Shinya Kanemura}
\email{kanemu@sci.u-toyama.ac.jp}
\affiliation{Department of Physics, University of Toyama, 3190 Gofuku,
Toyama 930-8555, Japan}
\author{Mariko Kikuchi}
\email{kikuchi@jodo.sci.u-toyama.ac.jp}
\affiliation{Department of Physics, University of Toyama, 3190 Gofuku,
Toyama 930-8555, Japan}
\author{Kei Yagyu}
\email{K.Yagyu@soton.ac.uk}
\affiliation{
School of Physics and Astronomy, University of Southampton, Southampton,
SO17 1BJ, United Kingdom}
\author{Hiroshi Yokoya}
\email{hyokoya@sci.u-toyama.ac.jp}
\affiliation{Department of Physics, University of Toyama, 3190 Gofuku,
Toyama 930-8555, Japan}

\begin{abstract}
In this Letter, we study the latest bound on the mass of doubly charged
 Higgs bosons, $H^{\pm\pm}$, assuming that they dominantly decay into a
 diboson.
The new bound is obtained by comparing the inclusive searches for
 events with a same-sign dilepton by the ATLAS Collaboration using the
 latest 20.3~fb$^{-1}$ data at the LHC 8~TeV run with theoretical
 prediction based on the Higgs triplet model with next-to-leading order
 QCD corrections.
We find that the lower mass bound on $H^{\pm\pm}$ is about 84~GeV.
\end{abstract}

\preprint{UT-HET 098}

\maketitle

%%%%%%%%%%%%%%%%%%%%%%%%%%%%%%%%%%%%%%%%%%%%%%%%%%%%%%%%%%%%%%%%%%%%%%%%%%%%%%%
\section{Introduction}

Recently, the ATLAS Collaboration has released the new results for
the inclusive searches for events with a same-sign dilepton by using
the 20.3~fb$^{-1}$ data at the 8~TeV run of the
LHC~\cite{ATLAS:2014kca}.
They improve the previous results based on the 4.7~fb$^{-1}$ data at the
7~TeV run~\cite{ATLAS:2012mn}.
From non-observation of any excess from the standard model (SM)
background, upper limits at the 95\% confidence level (CL) on the
fiducial cross section have been obtained for inclusive production of the
same-sign dilepton from the non-SM contribution.

One of the most interesting applications of these results is to obtain a
constraint on the parameter space for physics related to doubly charged
Higgs bosons $H^{\pm\pm}$. 
In various exotic models beyond the SM, $H^{\pm\pm}$ are predicted its
existence, e.g., in the left-right symmetric model~\cite{lr}, in models
with the type-II seesaw mechanism~\cite{typeII}, and in neutrino mass
models via quantum effects~\cite{radiative}. 
In this Letter, we focus on $H^{\pm\pm}$ in the Higgs triplet
model~(HTM)~\cite{typeII}, where the Higgs sector is composed of an
isospin doublet Higgs field with the hypercharge $Y=1/2$ and a triplet
field with $Y=1$. 
In the HTM, two decay modes are allowed for $H^{\pm\pm}$, i.e., decays
into the same-sign dilepton and the same-sign diboson.\footnote{%
If $H^{\pm\pm}$ are heavier than singly charged scalar bosons $H^\pm$,
$H^{\pm\pm}$ can also decay into $H^\pm W^\pm$~\cite{Aoki:2011pz}.}
If the same-sign dilepton decay is dominant, the most stringent lower
limit on the mass of $H^{\pm\pm}$ $(m_{H^{\pm\pm}})$ has been obtained
to be about 550 GeV~\cite{ATLAS:2014kca} at the LHC.

On the other hand, searches for $H^{\pm\pm}$ in the same-sign diboson
decay mode are of distinct importance.
The detection of interactions to weak gauge bosons can probe that
$H^{\pm\pm}$ come from Higgs fields with a non-trivial isospin charge.
When the same-sign diboson decay is dominant, the mass bound given in
the above is no longer applied.
In our previous publications~\cite{Kanemura:2013vxa,Kanemura:2014goa},
we have performed analyses to obtain the mass bound of $H^{\pm\pm}$
in the diboson decay scenario in the HTM.
By using the results on the inclusive searches for events with a same-sign
dilepton at the LHC~\cite{ATLAS:2012mn}, the obtained mass bound of
$H^{\pm\pm}$ has been
$m_{H^{\pm\pm}}\gtrsim60$~GeV~\cite{Kanemura:2014goa}.
In this Letter, we update our analysis based on the new data in
Ref.~\cite{ATLAS:2014kca}, and revise the mass bound of $H^{\pm\pm}$ in
the diboson decay scenario.

%%%%%%%%%%%%%%%%%%%%%%%%%%%%%%%%%%%%%%%%%%%%%%%%%%%%%%%%%%%%%%%%%%%%%%%%%%%%%%%
\section{ATLAS new results at the 8~TeV run}

In Ref.~\cite{ATLAS:2014kca}, the inclusive searches for events with
a same-sign dilepton have been performed by the ATLAS Collaboration by
using the full data set at the 8~TeV run of the LHC.
Events which contain a same-sign dilepton have been collected with the
selection cuts of 
(i)   $p_T>25$~GeV for the leading transverse momentum ($p_T$) lepton,
(ii)  $p_T>20$~GeV for the sub-leading $p_T$ lepton, 
(iii) $|\eta|<2.5$ for both leptons where $\eta$ represents the
pseudorapidity and (iv) an invariant mass cut of $M_{\ell\ell}>15$~GeV.
To reduce background from $Z$ boson decays, (v) events with
an opposite-sign same-flavor dilepton whose invariant mass satisfies
$|M_{\ell\ell}-m_Z|<10$~GeV are rejected.
In addition, in the $e^\pm e^\pm$ channel, (vi) events with a same-sign
dielectron in the mass range between 70~GeV and 110~GeV are vetoed to
use events in this region as a control sample to estimate the SM
background.
Total numbers of the collected events and invariant mass distributions are
in good agreement with the prediction by the SM, and therefore upper
limits on the cross section from the non-SM contribution are obtained
for the fiducial region defined above.

%%%%%%%%%%%%%%%%%%%%%%%%%%%%%%%%%%%%%%%%%%%%%%%%%%%%%%%%%%%%%%%%%%%%%%%%%%%%%%%
\section{Limit on $H^{\pm\pm}$ in the diboson decay scenario}

The experimental limits on the fiducial cross section can be compared
with the theoretical prediction calculated as 
\begin{align}
 \sigma_{\rm fid} = \sigma_{\rm tot}\cdot{\mathcal B}\cdot\epsilon_{A}, 
\end{align}
where $\sigma_{\rm tot}\cdot{\mathcal B}$ is (sum of) the total cross
section times branching ratio for the process giving the same-sign
dilepton signal from the new physics model, and $\epsilon_A$ is the factor
of efficiencies of the acceptance and kinematical cuts.
We evaluate the fiducial cross section for the process with the
same-sign dimuon, $\mu^\pm\mu^\pm$, in the final state via 
$H^{\pm\pm}\to W^{(*)\pm}W^{(*)\pm}$ in the HTM.
The other channels, such as $e^\pm e^\pm$ and $e^\pm\mu^\pm$ turn out to
give weaker bounds than the $\mu^\pm\mu^\pm$ channel. 
In the following discussion, we assume that the branching ratio of the
diboson decay mode is 100\%.\footnote{%
This scenario can be realized by taking the vacuum expectation value of
the triplet field to be larger than about $10^{-4}$
GeV~\cite{diboson}.}
The branching ratio for the $H^{\pm\pm}\to
W^{(*)\pm}W^{(*)\pm}\to\mu^{\pm}\mu^{\pm}\nu\nu$ channel is explained in
details in Ref.~\cite{Kanemura:2014goa}.

The dominant production processes of $H^{\pm\pm}$ at the LHC are
$(a)$ $pp\to H^{++}H^{--}$,
$(b)$ $pp\to H^{++}H^{-}$, and
$(c)$ $pp\to H^{+}H^{--}$,
where $H^\pm$ are the singly charged Higgs bosons which is also
introduced in the HTM.
The total cross sections for these processes have been calculated up to
the next-to-leading order (NLO) in perturbative QCD~\cite{Kfactor}.
Numerical predictions at the LHC with various collision energies can be
found in Ref.~\cite{Kanemura:2014goa}.
We assume that the mass of $H^\pm$ is the same as that of $H^{\pm\pm}$
for simplicity.

In this Letter, efficiencies for the acceptance and kinematical
cuts are estimated by using {\tt MadGraph5}~\cite{MG5} for each
production process at the parton level in the leading order.
Because we consider only inclusive production of a pair of same-sign
muons, and do not count the other particles, the cuts (v) and (vi)
explained in the last section are omitted.
In Table~\ref{tab}, we summarize the total cross sections, branching
ratio and the efficiencies for $m_{H^{\pm\pm}}=50$~GeV to 100~GeV.
By combining them, the fiducial cross section for the inclusive
$\mu^\pm\mu^\pm$ production is calculated as
\begin{align}
 \sigma_{\rm fid}(\mu^\pm\mu^\pm) = &\left[
\sigma_a\cdot\left\{2\epsilon_a-\epsilon_a^2{\mathcal
 B}_{\mu\mu}\right\}
+ \sigma_b\cdot\epsilon_b
+ \sigma_c\cdot\epsilon_c
\right]\cdot{\mathcal B}_{\mu\mu},
\end{align}
where $\sigma$, $\epsilon$ are the total cross sections, efficiencies
for the processes $(a)$, $(b)$ and $(c)$, respectively, and ${\mathcal
B}_{\mu\mu}={\mathcal B}(H^{\pm\pm}\to\mu^\pm\mu^\pm\nu\nu)$. 
The results for the fiducial cross sections are also summarized in
Table~\ref{tab}. 

\begin{table*}[bth]
 \begin{tabular}{l|rrrrrrrc}
  \hline
  $m_{H^{\pm\pm}}$&  50 & 60 & 70 & 80 & 90 & 100 &  [GeV] \\
  \hline \hline
  $\sigma^{\rm NLO}_{\rm tot}(pp\to H^{++}H^{--})$ & 8.52 & 3.57
	  & 1.93 & 1.16 & 0.744 & 0.501 & [pb] \\
  $\sigma^{\rm NLO}_{\rm tot}(pp\to H^{++}H^{-})$ [$m_{H^{\pm}} =
  m_{H^{\pm\pm}}$] & 10.6 & 4.47 & 2.36 & 1.40 & 0.891 & 0.598 & [pb]
			      \\
  $\sigma^{\rm NLO}_{\rm tot}(pp\to H^{+}H^{--})$ [$m_{H^{\pm}} =
  m_{H^{\pm\pm}}$] & 6.71 & 2.73 & 1.40 & 0.803 & 0.498 & 0.326 & [pb]
			      \\ \hline
  ${\mathcal B}(H^{\pm\pm}\to\mu^\pm\mu^\pm\nu\nu)$ & 2.22 & 2.21 & 2.19
	      & 2.16 & 1.98 & 1.61 & [\%] \\ \hline
  $\epsilon_{A}(pp\to H^{++}H^{--})$ & 5.1 & 9.9 & 16. & 21. & 23. &
			  23. & [\%] \\ 
  $\epsilon_{A}(pp\to H^{++}H^{-})$ & 4.9 & 9.9 & 15. & 21. & 22. &
			  23. & [\%] \\ 
  $\epsilon_{A}(pp\to H^{+}H^{--})$ & 4.7 & 9.7 & 15. & 21. & 23. &
			  22. & [\%] \\ \hline \hline
  $\sigma_{\rm fid}(pp\to\mu^\pm\mu^\pm+X)$ [$m_{H^{\pm}} =
  m_{H^{\pm\pm}}$] & 37.7 & 31.2 & 26.2 & 20.2 & 12.8 & 6.98 & [fb] \\
  \hline
 \end{tabular}
 \caption{Table of the total cross sections~\cite{Kanemura:2014goa} for
 $H^{++}H^{--}$, $H^{++}H^{-}$ and $H^{+}H^{--}$ processes, branching
 ratio of $H^{\pm\pm}$ into a same-sign dimuon~\cite{Kanemura:2014goa},
 and the acceptance and cut efficiencies for $\mu^\pm\mu^\pm$ searches
 at the LHC with 8~TeV for $m_{H^{\pm\pm}}=50$~GeV to 100~GeV.
 Efficiencies include acceptance cuts for $p_T$, $\eta$ of muons, and
 the invariant mass cut $M_{\mu\mu}>15$~GeV. 
 The resulting fiducial cross section is also listed. }\label{tab} 
\end{table*}

\begin{figure}[t]
 \begin{center}
  \includegraphics[width=100mm]{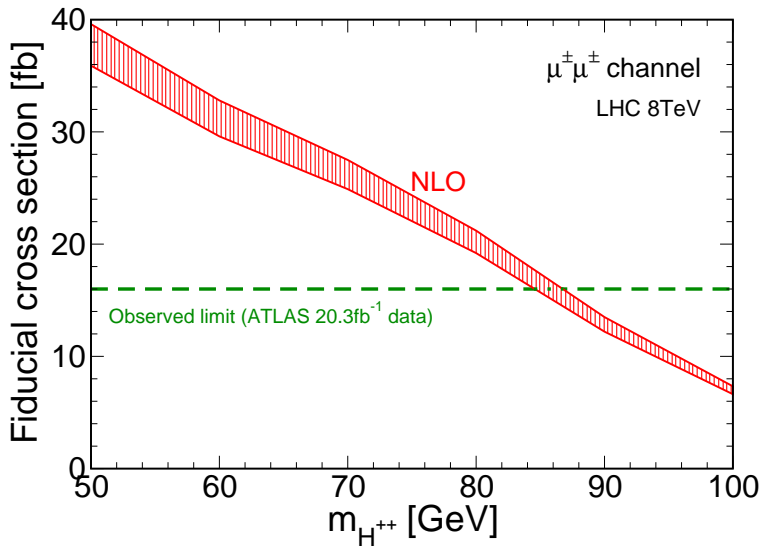}
  \caption{The fiducial cross section for the $\mu^\pm\mu^\pm$ channel
  at the LHC 8~TeV collision as a function of $m_{H^{\pm\pm}}$.
  The green dashed horizontal line shows the 95\% CL upper limit from
  the ATLAS data of the integrated luminosity to be
  20.3~fb$^{-1}$~\cite{ATLAS:2014kca}. 
  Red shaded band shows the NLO prediction with 5\% uncertainty.
  Details can be found in Table~\ref{tab}. 
  }
  \label{fig}
 \end{center}
\end{figure}

Now, we are ready to compare the fiducial cross sections for the
inclusive $\mu^\pm\mu^\pm$ production via the diboson decay of
$H^{\pm\pm}$ at the LHC. 
In Fig.~\ref{fig}, the fiducial cross section for the $\mu^\pm\mu^\pm$
events is plotted as a function of $m_{H^{\pm\pm}}$. 
Red band shows the NLO prediction, where its width indicates 5\%
uncertainty from scale variation and errors from parton distribution
functions~\cite{Lai:2010vv}. 
The green dashed horizontal line shows the 95\% CL upper limit obtained by
the ATLAS Collaboration,
\begin{align}
 \sigma^{\rm fid}_{95}(\mu^\pm\mu^\pm,M_{\mu\mu}>15~{\rm GeV})
 = 16~{\rm [fb]}.
\end{align}
By comparing them, we find that doubly charged Higgs bosons with
$m_{H^{\pm\pm}}\lesssim84$~GeV are excluded in the diboson decay
scenario.
For the reference, the experimental limits for the other decay channels
are reported as
$\sigma^{\rm fid}_{95}(e^\pm e^\pm,M_{ee}>15~{\rm GeV}) = 32~{\rm [fb]}$
and $\sigma^{\rm fid}_{95}(e^\pm\mu^\pm,M_{e\mu}>15~{\rm GeV}) = 29~{\rm
[fb]}$~\cite{ATLAS:2014kca}, while theoretical estimations for these
channels are comparable with the $\mu^\pm\mu^\pm$ channel in the mass
range of $m_{H^{\pm\pm}}\lesssim90$~GeV~\cite{Kanemura:2014goa}.
Thus, the limits by these channels are negligible.

\section{Conclusion}

We have studied the latest mass bound on the doubly charged
Higgs bosons in the diboson decay scenario in the HTM.
The new limit has been obtained by comparing the inclusive searches of
events with a same-sign dilepton by the ATLAS Collaboration using the
latest 20.3~fb$^{-1}$ data set at the LHC 8~TeV
run~\cite{ATLAS:2014kca} with theoretical prediction which includes the
production cross section with NLO QCD corrections, branching ratio with
interference effects, and efficiencies for the acceptance and
kinematical cuts~\cite{Kanemura:2014goa}. 
The lower bound has been revised to be $m_{H^{\pm\pm}}\gtrsim84$~GeV.

\section*{Acknowledgments}
We thank Koji Terashi for useful discussions.
This work was supported in part by Grant-in-Aid for Scientific Research,
Nos.\ 22244031, 23104006 and 24340046, JSPS, No.\ 25$\cdot$10031, and
JSPS postdoctoral fellowships for research abroad.

\end{document}